\documentclass[sigplan,screen,nonacm]{acmart}

\usepackage{booktabs}
\usepackage{graphicx}
\usepackage{microtype}
\usepackage{enumitem}
\usepackage{float}
\usepackage{xurl}
\graphicspath{{figures/}}

\title[Shift schema drift left]{Shift schema drift left: policy-aware compile-time contracts for typed JVM and Spark pipelines}

\author{\href{https://vitthalmirji.com/}{Vittal Mirji}}
\orcid{0009-0005-3376-0457}
\affiliation{
  \position{Staff Software Engineer}
  \country{ }
}
\email{vitthalmirji@gmail.com}
\AtBeginDocument{
  \hypersetup{
    pdfauthor={Vittal Mirji},
    pdftitle={Shift schema drift left: policy-aware compile-time contracts for typed JVM and Spark pipelines},
    pdfsubject={A Scala 3 and Spark mechanism paper on compile-time structural compatibility and sink-boundary runtime schema checks},
    pdfkeywords={Scala 3, Spark, macros, data contracts, schema drift, compile-time verification}
  }
}

\begin{document}

\begin{abstract}
Schema drift in data pipelines is often caught only when a job touches real data. Typed-Dataset layers close part of this gap but require wholesale adoption; table-level enforcement systems close another part but operate at write time against a stored schema. We present a small Scala~3 framework that occupies the seam: it proves producer-to-contract \emph{structural} compatibility under explicit policies at compile time, derives Spark schemas from the same contract types, and re-checks the actual \texttt{DataFrame} schema at the sink boundary before write. The artifact fuses the compile-time witness with a policy-aware runtime comparator that adds a nested-collection-optionality check Spark's built-in comparators omit and implements structural subset semantics for backward- and forward-compatible field sets. Evaluation covers compile-time proofs, runtime policy tests, builder-path end-to-end tests, and reproducible benchmarks on two environments. This is a narrow, honest mechanism artifact; the broader claim that compile-time structural contracts deliver measurable productivity or reliability in practice is stated as motivation and left for future work.
\end{abstract}


\maketitle
\pagestyle{plain}

\section{Introduction}

Schema drift is still found too late in many data systems. A producer renames a field, reorders a record, widens optionality, or changes the nullability of a nested collection, and the break often appears only when a job touches real data. Spark offers runtime schema comparators and schema-aware readers, but those checks still run after the program has been built and after an external boundary has already been crossed. A pipeline can therefore look typed in code while the key compatibility question is deferred to execution time~\cite{spark_datatype}.

Typed-Dataset layers and table-level schema enforcement each close part of this gap but leave a seam. Typed-Dataset libraries such as \texttt{frameless} and \texttt{iskra} require that raw Spark \texttt{DataFrame} code be replaced with a typed wrapper across the job, which raises adoption cost in practice~\cite{frameless,iskra}. Table-level schema enforcement such as Delta Lake and Apache Iceberg operates at write time against a stored schema rather than at compile time against a declared Scala type, and therefore still runs after job submission~\cite{delta_lake,apache_iceberg}. Neither layer currently provides compile-time proof that a declared producer type conforms to a declared contract type under an explicit policy before the job is submitted at all. That is the seam this paper's artifact targets. The value of closing it is argued here and left for future work to measure.

For typed JVM pipelines, much of that question is structural and available earlier. If the producer and sink contract are both represented as Scala types, then field names, nesting, collection shape, optionality, and defaulted fields are visible to the compiler. Earlier Scala derivation routes such as \texttt{shapeless} and Scala~3 \texttt{Mirror}-based derivation show adjacent implementation paths, but this artifact uses quoted reflection directly because it exposes \texttt{TypeRepr}, symbols, and type structure without introducing a separate generic wrapper layer~\cite{shapeless,scala3_reflection}. That gives us a concrete way to compute a normalized structural model at compile time and reject incompatible producer-to-contract pairs before a pipeline ships.

Compile time is not enough. Even if application code proves that a declared producer type conforms to a contract, the actual runtime schema can still drift at the data boundary. CSV, JSON, and other external inputs do not obey the Scala types used in pipeline code. The artifact in this paper therefore combines two checks: compile-time proof that the declared producer type conforms to the target contract under an explicit policy, and a runtime pin that validates the actual Spark \texttt{DataFrame} schema before write. The first shifts structural drift detection into compilation and CI; the second protects the external boundary where runtime data can still violate assumptions.

This paper presents a deliberately narrow framework for that combination. It derives normalized structural shapes from Scala~3 case classes, materializes compile-time evidence \texttt{SchemaConforms[Out, Contract, P]} under an explicit policy family, derives Spark schemas from the same contract types, and enforces a policy-aware runtime comparator at the sink boundary. Where Spark's documented comparators stop, the runtime layer adds a deeper check for nested collection optionality, which the artifact shows as a real drift source in tests~\cite{spark_datatype}.

The artifact is structural rather than semantic. It does not model business rules, temporal properties, cross-record invariants, or external schema-registry workflows. That boundary matters. Prior work on software contracts shows that contracts are useful in practice, but also that the space is much richer than boundary shape checking; temporal and effectful contracts, for example, are well beyond the scope of this artifact~\cite{contracts_in_practice,trace_contracts,effectful_software_contracts}.

Within that scope, this paper makes three contributions:

\begin{enumerate}[leftmargin=1.5em]
  \item A Scala~3 macro derivation of normalized structural shapes that rejects incompatible producer-to-contract pairs at compile time under an explicit, tested policy family, with path-rich drift diagnostics.
  \item A sink-boundary design that fuses that compile-time witness with a policy-aware Spark runtime comparator, including subset semantics for \texttt{Backward}/\texttt{Forward} and a nested-collection-optionality check that Spark's built-in comparators omit.
  \item A reproducible artifact demonstrating that this combination can be carried at low additional compile cost and at write-time runtime cost that is higher than Spark's built-in comparators but still within a single-digit microsecond per schema comparison.
\end{enumerate}

The paper is therefore best read as a mechanism paper with an honest artifact, not as a production-effectiveness paper. What is proven here is that a focused class of structural drift checks can move from runtime failure toward compile-time proof and CI while still retaining a runtime guard at the actual data boundary.

\section{Background and model}

\subsection{Structural contracts in this paper}

The term \emph{data contract} is overloaded. In some systems it includes business semantics, quality thresholds, temporal guarantees, ownership, and operational policy. In this paper, a contract is narrower: it is the structural shape expected at a checked sink boundary. That shape includes field names, field order where the chosen policy cares about order, nesting through product types, sequence and map structure, field optionality, nested collection optionality, and whether a contract field has a default value. It does not include domain constraints such as valid ranges, allowed enum values, uniqueness across records, or temporal properties across events.

This choice is deliberate. The artifact is strongest where the compiler and the Spark schema model have a direct structural correspondence. Scala~3 macros can inspect case-class structure at compile time, and Spark represents row schemas using \texttt{StructType}, \texttt{StructField}, \texttt{ArrayType}, and \texttt{MapType}~\cite{scala3_reflection,spark_structtype}. That makes a focused structural contract model practical to derive and compare from both sides of the pipeline.

We therefore use the following working definition:

\begin{quote}
A structural data contract is a typed description of the record shape expected at a checked boundary, together with a comparison policy that defines what counts as compatible producer-to-contract drift.
\end{quote}

This framing leaves room for richer contract systems in future work while keeping the current claims honest. Field-level optionality is intentionally ignored under all non-\texttt{Full} policies to match Spark's default field-nullability semantics; nested optionality inside arrays and map values is preserved and compared explicitly.

\subsection{Policy family}

Compatibility is not a single relation. Different boundaries need different answers to ``does this producer conform to that contract?'' The artifact makes that choice explicit through a small policy family:

\begin{itemize}[leftmargin=1.5em]
  \item \texttt{Exact} and \texttt{ExactUnorderedCI}: unordered matching by field name, case-insensitive, with no extras and no missing fields.
  \item \texttt{ExactOrdered}: ordered matching by field name, case-sensitive.
  \item \texttt{ExactOrderedCI}: ordered matching by field name, case-insensitive.
  \item \texttt{ExactByPosition}: by-position matching where field names are ignored.
  \item \texttt{Backward}: contract-by-name subset semantics. Producer extras are allowed; missing contract fields are allowed only when the contract field is optional or has a default value.
  \item \texttt{Forward}: producer-by-name subset semantics. Producer fields must all exist in the contract, while the contract may contain additional fields.
  \item \texttt{Full}: accept all structural combinations.
\end{itemize}

These policies separate intent from implementation. A sink can declare whether it wants exact matching, position-based matching, or subset-style compatibility with default relaxations. The compile-time and runtime layers then enforce that declared intent at their respective boundaries.
The subset-style policies are intentionally case-sensitive at both layers to avoid ambiguity in subset direction. A case-insensitive variant would be a reasonable extension, but it is not part of the current artifact.

\subsection{Why compile-time proof and runtime pin both exist}

Compile-time proof answers a question about declared program structure: given producer type \texttt{Out}, contract type \texttt{R}, and policy \texttt{P}, can the compiler prove that \texttt{Out} structurally conforms to \texttt{R} under \texttt{P}? That is a property of code. Runtime pinning answers a different question: does the actual Spark schema reaching the sink still satisfy the declared contract under the same policy? That is a property of runtime data.

The separation matters because data pipelines cross external boundaries. A typed source or transform can be declared against a case class, while the data actually read from files or tables may still drift independently. Compile-time proof cannot see those runtime schemas. Conversely, runtime validation alone comes too late to catch declared drift in the code that wires the pipeline. The artifact therefore treats the two checks as complementary rather than redundant. This split is deliberate: unlike designs that collapse structural validation into a single pass, the artifact separates the checks because they consume different inputs at different times, namely declared Scala types at compile time and external Spark schemas at the sink boundary.

Spark's documented comparator surface helps explain the runtime side. The three relevant comparison entry points are:

\begin{itemize}[leftmargin=1.5em]
  \item \texttt{equalsIgnoreCaseAndNullability} for unordered name-based comparison with case-insensitive resolution and ignored field nullability,
  \item \texttt{equalsStructurally} for by-position comparison, and
  \item \texttt{equalsStructurallyByName} for ordered-by-name comparison with a supplied name resolver~\cite{spark_datatype}.
\end{itemize}

These comparators are a useful baseline for exact-style runtime matching, but they are not enough for the artifact's full policy story. The artifact also needs subset semantics for \texttt{Backward} and \texttt{Forward}, and it needs nested collection optionality checks for arrays and map values because Spark's default comparators do not cover that drift.

\subsection{Relationship to broader contract work}

The model here is intentionally smaller than the broader software-contract literature. Empirical work shows that contracts matter in real software projects, while more recent work on effectful and trace-oriented contracts shows that the contract design space extends far beyond record shape compatibility~\cite{contracts_in_practice,effectful_software_contracts,trace_contracts}. The right comparison for this paper is therefore not ``have we solved contracts?'' but ``have we built and evidenced a useful compile-time plus runtime mechanism for a focused structural boundary problem?''

\section{Framework design}

\subsection{Overview}

The framework has one central design rule: the same contract type should drive both the compile-time proof path and the runtime sink pin. The compile-time path starts from Scala types and produces a structural witness. The runtime path starts from a Spark \texttt{DataFrame} schema and checks it against a \texttt{StructType} derived from the same contract type. The typed builder then fuses the two by requiring compile-time evidence at \texttt{addSink[R, P]} and by re-checking the actual schema before write.

That design is what prevents the framework from becoming only a macro demo or only a runtime checker. If the compile-time path is not tied to the actual sink boundary, the paper risks proving a property of declarations while ignoring the data path. If the runtime path is not tied back to the same contract type, the paper becomes a Spark schema-checking utility without a compile-time story. The artifact keeps those two sides connected through the contract type \texttt{R} and the chosen policy \texttt{P}.

\begin{figure}[H]
  \centering
  \includegraphics[width=\columnwidth]{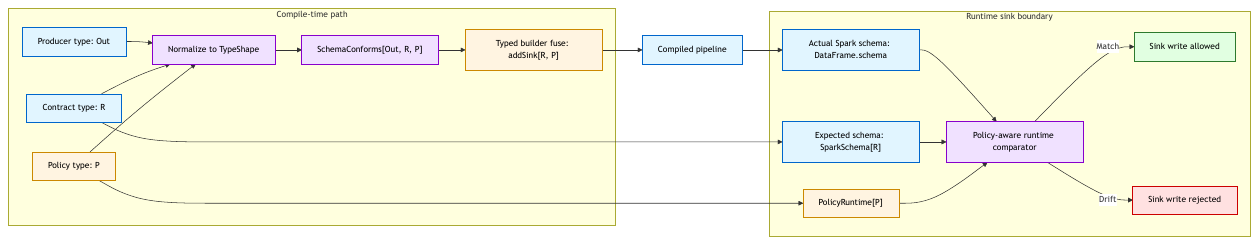}
  \caption{The same contract type \texttt{R} drives both compile-time proof and the runtime sink pin. Compile time rejects declared producer-to-contract drift in code. Runtime re-checks the actual Spark schema before write because external data boundaries can still violate the declared type path.}
  \Description{Architecture overview of the framework. The compile-time path derives a normalized shape from producer type Out, contract type R, and policy P, then materializes SchemaConforms evidence required by the typed sink builder. The runtime path derives SparkSchema from R, applies PolicyRuntime for P to the actual DataFrame schema, and either allows or rejects the sink write.}
  \label{fig:architecture}
\end{figure}

\subsection{Normalized structural model}

The compile-time side needs an internal representation that is small enough to reason about and expressive enough to cover the supported shapes. The artifact uses a normalized \texttt{TypeShape} model with primitive shapes, optional shapes, sequence shapes, map shapes with atomic keys, and struct shapes whose fields carry four pieces of information: field name, nested shape, whether the field has a default value, and whether the field is optional.

This representation is intentionally not user-facing. Its job is to normalize different Scala surface forms into one comparison model. Once two Scala types have been reduced to \texttt{TypeShape}, policy-specific comparison becomes a deterministic structural walk instead of ad hoc pattern matching over raw reflection nodes.

The important detail is that optionality is tracked in two places. Field-level optionality is part of a field's own metadata. Nested optionality inside sequences and maps is represented in the nested shape itself. That distinction is what later allows the runtime layer to preserve and compare nested collection optionality instead of flattening it away.

\subsection{Compile-time evidence materialization}

Scala~3 quoted reflection makes it possible to materialize compatibility evidence directly at the call site~\cite{scala3_reflection}. The artifact defines a type class \texttt{SchemaConforms[Out, Contract, P]}. An inline given macro derives that witness by inspecting \texttt{Out} and \texttt{Contract} through \texttt{quotes.reflect}, computing normalized shapes for both sides, comparing them under the selected \texttt{SchemaPolicy}, and either returning the witness or aborting compilation with a path-rich drift report.

The drift report is part of the design, not a side effect. If the framework only failed with a generic missing-given error, it would be much less useful in practice. The macro therefore records missing fields, extra fields, and mismatches with structural paths so that a failed compile points at the actual drift shape rather than only at the location where evidence was requested.

\subsection{Deriving runtime schemas from the same contract type}

The runtime path begins by deriving a Spark \texttt{StructType} from the contract case class. This derivation mirrors the supported compile-time shape family: primitives map to Spark atomic types, nested case classes map to nested \texttt{StructType}, sequence-like containers map to \texttt{ArrayType}, maps with atomic keys map to \texttt{MapType}, and \texttt{Option[T]} affects field or nested nullability. Default-valued contract fields are recorded in metadata so that runtime subset policies can make the same allowance that compile time makes for \texttt{Backward}.

The result is a runtime contract schema that is not manually duplicated. The same Scala contract type \texttt{R} is used for both compile-time comparison and runtime schema derivation. That keeps the sink boundary honest and avoids the common failure mode where one contract exists in typed code and a second, loosely synchronized version exists in runtime configuration.

\subsection{Policy-aware runtime comparison}

At runtime the framework selects a \texttt{PolicyRuntime[P]} instance. The exact-style policies mirror Spark's documented comparator families: unordered-by-name exact matching, ordered-by-name exact matching, and by-position exact matching~\cite{spark_datatype}. The artifact does not simply call Spark's built-in comparators and stop there. Instead, it implements a comparator that follows the same broad name-order directions while adding two behaviors that matter for the contract story: nested collection optionality is checked explicitly for arrays and map values, and \texttt{Backward} and \texttt{Forward} are implemented as real subset relations rather than being collapsed into equality.

This is the most important runtime design decision in the artifact. Without it, the framework would overclaim semantic parity between compile-time and runtime enforcement. With it, the runtime layer becomes a true boundary check for the policies the paper claims to support.

\begin{figure}[H]
  \centering
  \includegraphics[width=\columnwidth]{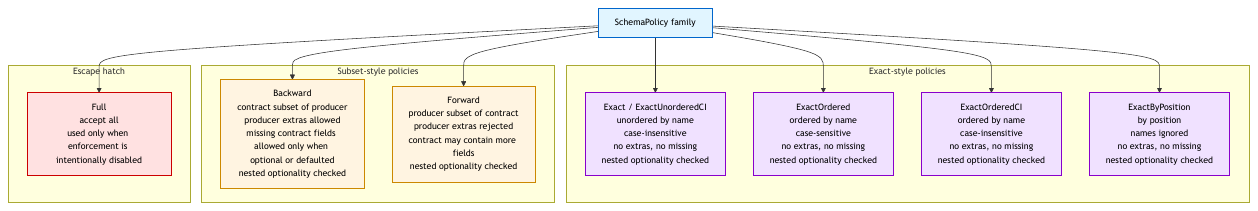}
  \caption{Policy family supported by the artifact. Exact-style policies differ on name and order sensitivity. Subset-style policies make direction explicit. Nested collection optionality is checked across all policies except \texttt{Full}.}
  \Description{Policy overview diagram. Exact and ExactUnorderedCI are unordered by name and case-insensitive. ExactOrdered and ExactOrderedCI preserve field order. ExactByPosition compares by position. Backward allows producer extras and optional or defaulted contract omissions. Forward allows the contract to contain more fields than the producer. Full accepts all combinations.}
  \label{fig:policy-family}
\end{figure}

Figure~\ref{fig:policy-family} summarizes the supported policy family at a glance. The exact-style policies differ on name and order sensitivity, while the subset-style policies make direction explicit.

\subsection{Typed builder fuse}

The user-facing enforcement point is the typed builder. A sink is added through \texttt{addSink[R, P]}, and that call requires compile-time evidence that the current producer type conforms to contract \texttt{R} under policy \texttt{P}. If evidence cannot be derived, the pipeline does not compile. If evidence is derived, the built pipeline still performs the runtime schema pin before write.

This builder encodes a lightweight typestate discipline~\cite{aldrich_typestate_2009}: the phantom type parameter tracks which capabilities the builder has acquired, and the type system prevents illegal transitions such as adding a sink before a source exists. This builder surface matters because it ties the proof obligation to the checked boundary itself. The compile-time witness is not a separate demonstration step that users can forget to wire in. It is part of the sink-construction path. The runtime pin is not an optional logging step after the fact. It is part of the write path. Mid-pipeline pins intentionally use the default unordered case-insensitive comparator to catch gross shape drift during reshaping; policy-aware enforcement is applied at the sink boundary only, where the declared contract type is tied to the actual write path. That fusion is what turns the artifact from a library of helpers into a coherent boundary-enforcement design.

\section{Artifact and evaluation}

\subsection{Artifact scope}

We evaluate the clean reference repository rather than the larger framework effort that motivated it. The repository is intentionally small. It contains the compile-time policy model, the Scala~3 macro derivation for \texttt{SchemaConforms}, Spark schema derivation from contract types, a policy-aware runtime comparator, a typed builder path that enforces sink checks, and a reproducible benchmark harness.\footnote{A practitioner-oriented tutorial exposition of the same mechanism appears separately at \url{https://rockthejvm.com/articles/compile-time-data-contracts-scala-3}.} This is a mechanism evaluation: the question is whether the repository proves the structural properties it claims, not whether it already delivers industrial-scale outcomes.

Accordingly, the evaluation asks four narrow questions: whether compile-time witness generation accepts and rejects the intended structural cases; whether the runtime layer catches drift that declared Scala types cannot rule out, including subset-policy behavior and nested collection optionality drift; whether the typed builder path enforces both layers at the actual sink boundary; and whether the mechanism overhead is measurable and reproducible across the two saved environments.

\subsection{Compile-time and runtime coverage}

Compile-time coverage is centered on direct witness generation and compile-fail behavior. The tests exercise positive conformance for exact-style, backward-compatible, and forward-compatible policies, and rejection paths for reordering, missing required fields, and nested optionality drift. The covered shape family includes nested case classes, sequences, maps, optional fields, and defaulted contract fields.

The runtime layer is evaluated separately because it has a different job. The compile-time witness proves compatibility for declared Scala types. The runtime layer validates the actual Spark schema that reaches the sink. The runtime tests therefore ask whether the exact-style pin rejects nested collection optionality drift that Spark ignores by default, whether \texttt{Backward} and \texttt{Forward} behave as true subset relations at runtime, and whether contract-derived metadata allows missing defaulted fields where the policy permits them.

\subsection{Builder-path end-to-end checks}

The builder-path tests are the most operational part of the evaluation. They show that the paper is not only about a macro witness in isolation. \texttt{PipelineBuilder.addSink[R, P]} requires compile-time evidence for the current producer type, and the built pipeline still validates the actual \texttt{DataFrame} schema before write. The end-to-end checks cover a compile-time rejection at the sink, a green path where both the declared type and runtime schema satisfy the sink policy, a red path where the runtime schema violates the policy after compile time has succeeded, and no-transform subset-policy paths for \texttt{Backward} and \texttt{Forward}. This is the closest point in the artifact to the intended application boundary, so it is the strongest evidence that the two enforcement layers are wired together rather than merely coexisting in the same repository.

\subsection{Benchmark harness}

The benchmark harness is intentionally small. It measures compile-time witness-generation overhead on synthetic but representative schemas and runtime comparator overhead on nested \texttt{StructType} comparisons. It does not claim end-to-end Spark job performance.

The saved snapshots show compile-time deltas of \texttt{+0.270s}, \texttt{+0.397s}, and \texttt{+0.513s} on local \texttt{macOS arm64} for \texttt{10}, \texttt{25}, and \texttt{50} schema pairs. On GitHub-hosted \texttt{Ubuntu x86\_64}, the same sizes measure \texttt{+0.847s}, \texttt{+1.000s}, and \texttt{+1.880s}. Runtime unordered exact matching stays in the low-microsecond range per schema comparison on both environments, while exact-by-position matching stays below microsecond scale in both saved runs. The custom unordered comparator is roughly \texttt{17--25$\times$} slower than Spark's built-in \texttt{equalsIgnoreCaseAndNullability} baseline in the saved runs, but that cost remains acceptable for this artifact because the check runs once per sink write, not per row.

The right interpretation is limited but useful: the harness reproduces across two environments and shows that the mechanism is measurable and stable enough to package as artifact evidence. The wrong interpretation would be a broad claim about low overhead across machines or organizations. Table~\ref{tab:benchmarks} therefore reports bounded artifact evidence, not a general performance baseline.

\begin{table}[t]
  \caption{Saved benchmark snapshots for the artifact head. These numbers show reproducible harness output, not a broad cross-machine performance baseline.}
  \Description{Two-part benchmark table. The left half reports compile-time deltas for 10, 25, and 50 schema pairs on local macOS arm64 and GitHub-hosted Ubuntu x86_64. The right half reports runtime comparator averages for by-position matching, unordered exact matching, and two Spark baseline comparators on the same two environments.}
  \label{tab:benchmarks}
  \centering
  \scriptsize
  \textbf{Compile-time overhead}

  \begin{tabular}{lrrrr}
    \toprule
    Pairs & Local (s) & Local (\%) & Ubuntu (s) & Ubuntu (\%) \\
    \midrule
    10 & 0.270 & 11.8 & 0.847 & 12.6 \\
    25 & 0.397 & 13.5 & 1.000 & 11.1 \\
    50 & 0.513 & 13.9 & 1.880 & 16.6 \\
    \bottomrule
  \end{tabular}

  \vspace{0.6em}
  \textbf{Runtime comparator averages}

  \begin{tabular}{lrr}
    \toprule
    Benchmark & Local (ns) & Ubuntu (ns) \\
    \midrule
    By-position & 116.82 & 180.55 \\
    Unordered exact & 4736.41 & 8149.74 \\
    Spark ignore-case & 278.92 & 331.42 \\
    Spark structural & 332.13 & 380.36 \\
    \bottomrule
  \end{tabular}
\end{table}

\section{Related work}

This paper sits at the intersection of software contracts, static-versus-dynamic checking, and typed data-pipeline boundaries. The closest primary sources for the implementation itself are the Scala~3 reflection documentation and the Spark schema-comparator APIs, because the artifact is built directly on those mechanisms~\cite{scala3_reflection,spark_datatype}. These are mechanism sources, not related-work contributions in the research sense.

The closest implementation-level neighbors are typed-Dataset libraries. \texttt{frameless} and \texttt{iskra} provide typed Spark layers that carry schema information through Dataset operations, but they do so by asking the user to replace raw \texttt{DataFrame} code with typed wrappers across the job~\cite{frameless,iskra}. This artifact targets a narrower seam: producer-to-contract conformance is required only at the checked boundary, and the rest of the job may remain in raw Spark code.

Record-to-record derivation libraries solve a different adjacent problem. Chimney, and earlier Scala generic-derivation work built on \texttt{shapeless}, derive field-to-field transformations with rich compile-time diagnostics~\cite{chimney,shapeless}. That macro technique is close in spirit to the compile-time side of this artifact, but the goal is different: Chimney proves transformation existence, whereas this artifact proves structural conformance under an explicit policy family and couples that witness to a Spark sink-boundary pin.

Table-level schema enforcement is also close, but at a different layer. Delta Lake and Apache Iceberg validate writes against stored table metadata and support schema evolution at that boundary~\cite{delta_lake,apache_iceberg}. Those systems are valuable runtime seatbelts, but they operate after job submission and against stored schemas rather than against declared Scala producer and contract types. The artifact here is therefore complementary rather than competing.

The policy names \texttt{Backward} and \texttt{Forward} deliberately reuse vocabulary that is familiar from Avro and schema-registry compatibility systems~\cite{avro_spec,avro_compatibility}. In this paper, however, the relation is structural over Scala product types and Spark schemas. It is not a value-level compatibility proof for serialized messages or registered schemas.

The broader contract literature provides the right contrast. Empirical work such as \emph{Contracts in Practice} shows that contract disciplines remain relevant in real software, which supports the motivation for making contract intent explicit~\cite{contracts_in_practice}. More expressive systems such as \emph{Effectful Software Contracts} and \emph{Trace Contracts} show how much richer the contract space becomes when effects, histories, or traces are part of the property being checked~\cite{effectful_software_contracts,trace_contracts}. By comparison, this artifact is deliberately narrower: it checks structural compatibility at a typed boundary.

There is also a useful static-versus-dynamic comparison point. \emph{Dynamic Contract Analysis for Parallel Programming Models} treats dynamic contracts as complementary to static analysis rather than as a complete replacement~\cite{dynamic_contract_analysis}. That lesson aligns with this paper's design choice: compile-time proof and runtime checking do different jobs and should be combined at the boundary instead of being presented as competitors.

Finally, work outside the JVM ecosystem helps position the contribution. P3317R0 argues for contracts that can often be resolved at compile time so that runtime cost can be reduced or removed~\cite{compile_time_resolved_contracts}. The artifact here is not a direct analogue of that C++ proposal, but it shares the same directional idea: when part of a contract question is structural and visible to the compiler, deferring everything to runtime is unnecessary.

The novelty claim of this paper is therefore narrow. It is not a new general theory of contracts. It is a small, evidenced combination of compile-time structural proof and runtime boundary checking for a typed Scala and Spark setting.

\section{Limitations}

The first limitation is scope. The artifact checks structural compatibility, not business semantics. It can prove that two typed schemas match under a policy, but it does not know whether an \texttt{age} field must be positive, whether a \texttt{currency} field must come from an approved domain, or whether two records must satisfy a temporal or cross-row invariant. Those are different kinds of contract problems.

The second limitation is supported-shape breadth. The artifact intentionally focuses on product-like case-class schemas, nested products, sequence-like containers, maps with atomic keys, optional fields, nested collection optionality, and defaulted contract fields. That is enough for the mechanism paper, but it is not a full schema system for all Spark or Scala data representations.

The third limitation is ecosystem reach. There is no external schema-registry integration, no industrial deployment pack in the clean repo, and no measured user-study evidence about productivity or usability. The builder API may be teachable, but the paper should not convert that into a measured productivity claim.

The fourth limitation is evaluation breadth. The benchmark results come from two saved snapshots on one local machine and one GitHub-hosted Ubuntu runner. They show reproducibility of the harness, not a stable cross-machine baseline. The runtime numbers are micro-bench results for schema comparison, not end-to-end Spark job measurements.

The fifth limitation is value-add evidence. The paper argues that boundary-only compile-time proof may lower adoption cost relative to a typed-Dataset rewrite and may catch drift earlier than write-time schema enforcement, but it does not measure those claims. A follow-up evaluation should compare time-to-first-drift-catch in CI, drift-catches-per-month in a deployed pipeline, and adoption-cost delta versus a typed-Dataset rewrite or a \texttt{DataFrame} plus runtime-only pin.

The final limitation is the role of FlowForge. FlowForge informed the motivation, application-side realism, and overclaim analysis, but it does not serve as proof for the clean artifact's closed claims. That separation is a strength for honesty, but it also means the paper stops short of industrial-effectiveness claims for now.

\section{Conclusion}

This paper presents a focused mechanism for shifting structural contract drift detection left in typed JVM and Spark pipelines. The artifact proves compile-time structural compatibility under an explicit policy family, derives runtime schemas from the same contract types, and checks the actual Spark schema at the sink boundary before write.

The contribution is narrow by design: a bounded class of structural mismatches can be rejected at compile time and re-checked at runtime where external data still matters. The artifact does not claim semantic contracts, industrial metrics, or broad performance generalization.

That is the main result. Compile-time proof and runtime boundary checking belong together, and this artifact shows a small honest way to combine them while leaving a clean base for richer contract models and stronger industrial evidence.

\bibliographystyle{ACM-Reference-Format}
\bibliography{references}

\end{document}